\documentclass[12pt,epsfig]{article}
\usepackage{amsmath}
\usepackage{amssymb}
\usepackage{graphicx}
\usepackage{epsfig}
\newcommand{\bea}{\begin{eqnarray}}

\newcommand{\eea}{\end{eqnarray}}

\def\bes{\begin{eqnarray}}
 \def\ees{\end{eqnarray}}
\def\be{\begin{equation}}
\def\ee{\end{equation}}
\def\bs{\begin{subequations}}
\def\es{\end{subequations}}
\newcommand{\een}{\end{subequations}}
\newcommand{\ben}{\begin{subequations}}
\newcommand{\beq}{\begin{eqalignno}}
\newcommand{\eeq}{\end{eqalignno}}

\def\pit{{\tilde{\pi}}}

 \def\dpi{{\delta\pi}}

 \def\ex{\epsilon}
 \def\kx{\kappa}
 \def\Lx{\Lambda}
 \def\lx{\lambda}

 \def\kb{{\bar{\kappa}}}

\begin{document}

\flushright{CERN-PH-TH/2013-286}

\vspace{1.5cm}
\begin{center}
{ \Large \bf
Quantum corrections in Galileon theories}
\\
\vspace{1.5cm}
{\Large 
Nikolaos Brouzakis$^1$, A.~Codello$^2$, Nikolaos Tetradis$^{1,3}$ and O.~Zanusso$^{4}$ 
} 
\\
\vspace{0.5cm}
{\it
$^1$ Department of Physics, University of Athens,
Zographou 157 84, Greece
\\
$^2$ SISSA, Via Bonomea 265, 34136 Trieste, Italy
\\
$^3$ Department of Physics, CERN - Theory Division, CH-1211 Geneva 23, Switzerland
\\
$^4$ Radboud University Nijmegen, Institute for Mathematics, Astrophysics and Particle Physics,
 Heyendaalseweg 135, 6525 AJ Nijmegen, The Netherlands
} 
\end{center}
\vspace{3cm}
\abstract{
We calculate the one-loop quantum corrections in the cubic Galileon theory, using cutoff regularization. We
confirm the expected form of the one-loop effective action and that the couplings of the Galileon theory do not get
renormalized. However, new terms, not included in the tree-level action, are induced by quantum corrections.
We also consider the one-loop corrections in an effective brane theory, which belongs to the 
Horndeski or generalized Galileon class. We find that new terms are generated by quantum corrections, while the 
tree-level couplings are also renormalized. We conclude 
that the structure of the generalized 
Galileon theories is altered by quantum corrections more radically than that of the Galileon theory.

}
\newpage

\section{Introduction}\label{intro}

The Galileon theory describes the dynamics of the scalar mode that 
survives in the decoupling limit of the DGP model \cite{dgp}. The action contains a higher-derivative
term, cubic in the field $\pi(x)$, with a dimensionful coupling that sets the scale $\Lx$ at which the theory becomes strongly
coupled. The action is invariant under the Galilean transformation
$\pi(x)\to \pi(x)+b_\mu x^\mu+c$, up to surface terms.  Additional terms can also be present, but their number is
limited by the additional requirement that the theory does not contain ghost degrees of 
freedom  \cite{galileon}. Typically, the presence of a ghost 
is associated with an equation of motion that contains field derivatives higher than the second. The structure of
the Galileon theory guarantees that the equation of motion is a second-order partial differential
equation. This property can be preserved within a more general class of theories that do not possess the
Galilean symmetry. These were constructed some time ago \cite{horndeski} and rediscovered recently \cite{genegal}. 
They are characterized as  generalized Galileon theories.

The absence of higher-than-second derivatives in the equation of motion is 
a property not protected by some underlying symmetry. The Galilean symmetry of the Galileon theory 
reduces the number of allowed terms in the action, but the absence of derivatives
higher than the second is an independent requirement. 
It is natural, therefore, to question the consistency of the Galileon theory at the quantum level. 
An interesting property of this theory and some of its generalizations is  
the absence of perturbative renormalization of the couplings appearing in the tree-level action \cite{quantum1, nonrenorm,multifield}. 
This conclusion does not
exclude the possible emergence of terms not contained in the action of the 
Galileon theory. As first observed in \cite{quantum1}, such terms can be induced through quantum corrections. 
It was argued, however, that they are suppressed in certain regimes of physical interest, such as 
the scales at which the Vainshtein mechanism \cite{vainshtein} operates.

If a momentum cutoff is used, of the order of the fundamental scale $\Lx$ of the theory, the structure of the 
one-loop effective action of the Galileon theory is, schematically, \cite{quantum1,nonrenorm,multifield}
\be
\Gamma \sim \int d^4x \sum_m \left[ 
\Lx^4+\Lx^2\partial^2+\partial^4 \log \left(\frac{\partial^2}{\Lx^2} \right)
\right] \left( \frac{\partial^2 \pi}{\Lx^3} \right)^m.
\label{schem} \ee 
A calculation of the one-loop corrections in the cubic Galileon theory was
performed in \cite{quantum2} using dimensional regularization. It was found that the first correction 
that is quadratic in the field corresponds to a term $\sim\pi \Box^4 \pi$ in the action.
The two results are consistent, because the  
quartic and quadratic divergences within the bracket in eq.~(\ref{schem}) for $\Lx\to \infty$ 
are not visible through dimensional regularization. 
For $m=2$ the logarithmic correction corresponds to a term  $\sim\pi \Box^4 \pi$, in agreement with ref. \cite{quantum2}.

The first aim of our work is to confirm eq. (\ref{schem}) through an approach that employs a momentum cutoff, in order 
to reproduce all the terms for $m=2$. We perform the calculation for an arbitrary dimensionality $d$ of
space in order to gain intuition on the structure of the cubic Galileon theory in other dimensions. 
The main part of our work is
devoted to the calculation of quantum corrections for generalized Galileon theories. As their general structure is very complicated,
we concentrate on a theory of geometric origin: the theory of a brane embedded in flat space with one extra dimension, in the static gauge. 
The number of invariants in the action of this theory is limited, so that a one-loop calculation is feasible. In this way we obtain 
intuition on the effect of quantum corrections on the theories of the Horndeski class.

In the following section we discuss the one-loop quantum corrections for the cubic Galileon theory. In section \ref{brane} we
perform a similar calculation for the brane theory. Some technical aspects of this calculation are summarized in the appendix.
In section \ref{conclusions} we present our conclusions.

\section{Quantum corrections in the cubic Galileon theory}\label{cubicgalileon}

It is instructive to consider first the one-loop corrections within the cubic Galileon theory. 
The tree-level action in Euclidean $d$-dimensional space is  
\be
S_0=\int d^dx \left\{ \frac{1}{2} (\partial \pi)^2 -\frac{\nu_0}{2} (\partial \pi)^2 \Box \pi \right\},
\label{cubic} \ee 
where we have assumed that the field is canonically normalized.
Because of the large number of possible 
terms in the effective action, we focus our calculation on higher-derivative terms quadratic in the field. 

For the calculation of the effective action we consider a field fluctuation $\dpi$ around the background $\pi$,
and determine the part of the 
tree-level action quadratic in $\dpi$. Through some partial integrations it can be cast in the form
\be
S^{quad}_{0}=\int d^dx \left\{
-\frac{1}{2} \dpi \Box \dpi+\frac{\nu_0}{2} \dpi \left[ 
2(\Box \pi) \Box \dpi -2(\partial^\mu\partial^\nu\pi)\partial_\mu \partial_\nu \dpi
\right] \right\}.
\label{quad} \ee
We define the operators 
$K=- \Box$,
$\Sigma_1=2\nu_0 (\Box\pi)\, \Box$, 
$\Sigma_2=-2\nu_0 (\partial_\mu \partial_\nu \pi)\, \partial^\mu\partial^\nu$.
The one-loop correction to the effective action is 
\be S_1=\frac{1}{2}{\rm tr}\, \log \left(K+\Sigma_1+\Sigma_2 \right)=
\frac{1}{2}{\rm tr}\, \log \left(1+\Sigma_1 K^{-1}+\Sigma_2 K^{-1}\right) +{\cal N}.
\label{s1} \ee
The factor ${\cal N}=({\rm tr}\, \log K)/2$ contributes only to the vacuum energy and we neglect it in the following.
The expansion of the logarithm generates terms in the effective action that involve various powers of $\pi$.
As each of the operators $\Sigma_1$, $\Sigma_2$ involves only one power of $\pi$,  the quadratic part of the 
effective action is generated by the trace of $\Sigma_{1,2}K^{-1}\Sigma_{1,2}K^{-1}$. 

Using the Fourier transform $\pi(x)=\int d^dk\,\exp(i kx)\pit(k)$, 
we find
\be
{\rm tr} \left(\Sigma_{1}K^{-1}\Sigma_{1}K^{-1} \right)
=4\nu_0^2(2\pi)^d \int d^dk\,  k^4\pit(k)\pit(-k)\int\frac{d^dp}{(2\pi)^d}. 
\label{c11} \ee
Similarly, 
\be
{\rm tr} \left(\Sigma_{1}K^{-1}\Sigma_{2}K^{-1} \right)={\rm tr} \left(\Sigma_{2}K^{-1}\Sigma_1K^{-1} \right)=
-4\nu_0^2 (2\pi)^d \int d^dk\, k^4\pit(k)\pit(-k)\frac{1}{d}\int\frac{d^dp}{(2\pi)^d},
\label{c12} \ee
where we shifted the loop momenta by a constant and used the replacement $p^\mu p^\nu \to \eta^{\mu\nu}p^2/d$ 
within the loop integral. Finally,
\be
{\rm tr} \left(\Sigma_{2}K^{-1}\Sigma_{2}K^{-1} \right)=
4\nu_0^2 (2\pi)^d \int d^dk\, k_\mu k_\nu k_\rho k_\sigma \pit(k)\pit(-k)\int\frac{d^dp}{(2\pi)^d}
\frac{p^\mu p^\nu(p^\rho+k^\rho)(p^\sigma+k^\sigma)}{p^2(p^2+k^2)}.
\label{c22} \ee
We are interested in the UV divergences of the theory. The IR behavior can be determined precisely, as the $p$-integrals are IR finite
for nonzero external momenta $k$. On the other hand, we concentrate on the UV regime in the following. 
The UV divergences are more easily visible if we  
expand the integrant of eq. (\ref{c22}) in powers of $k$. 
The resulting expressions assume the presence of an IR cutoff of the order of the external momenta $k$.
The 
 integrals in eq. (\ref{c22}) are then evaluated easily after the replacements
\begin{eqnarray}
p^\mu p^\nu p^\rho p^\sigma &\to& \frac{p^4}{d(d+2)}
\left(\eta^{\mu\nu}\eta^{\rho\sigma}+\eta^{\mu\rho}\eta^{\nu\sigma}+\eta^{\mu\sigma}\eta^{\rho\nu}  \right)
\label{perm4} \\
p^\mu p^\nu p^\rho p^\sigma p^\kappa p^\lambda &\to& \frac{p^6}{d(d+2)(d+4)}
\left(\eta^{\mu\nu}\eta^{\rho\sigma}\eta^{\kx\lx}+{\rm 14\, permutations}  \right)
\label{perm6} \\
p^\mu p^\nu p^\rho p^\sigma p^\kappa p^\lambda p^\xi p^\tau&\to& \frac{p^8}{d(d+2)(d+4)(d+6)}
\left(\eta^{\mu\nu}\eta^{\rho\sigma}\eta^{\kx\lx}\eta^{\xi\tau}+{\rm 105\, permutations}  \right).
\label{perm8} \end{eqnarray}
We find that, up to order $k^8$,
\begin{eqnarray}
&&{\rm tr}\left(\Sigma_{2}K^{-1}\Sigma_{2}K^{-1} \right)=
4\nu_0^2(2\pi)^d \int d^dk\,  \pit(k)\pit(-k) \Bigg\{
\frac{3}{d(d+2)} k^4 \int\frac{d^dp}{(2\pi)^d} \Bigg.
\nonumber \\
 &+&\Bigg. \frac{(d-8)(d-1)}{d(d+2)(d+4)} k^6  \int\frac{d^dp}{(2\pi)^d}\frac{1}{p^2}
-\frac{(d-24)(d-2)(d-1)}{d(d+2)(d+4)(d+6)} k^8  \int\frac{d^dp}{(2\pi)^d}\frac{1}{p^4} \Bigg\}.
\label{c22e} \end{eqnarray}

Putting everything together, we obtain in position space
\begin{eqnarray}
&&S_1=\nu_0^2
\int d^dx\,  \pi(x) \Bigg\{
-\frac{d^2-1}{d(d+2)} \left( \int\frac{d^dp}{(2\pi)^d} \right)  \Box^2 \Bigg.
\nonumber \\
 &+&\Bigg. \frac{(d-8)(d-1)}{d(d+2)(d+4)} \left( \int\frac{d^dp}{(2\pi)^d}\frac{1}{p^2} \right) \Box^3 
+\frac{(d-24)(d-2)(d-1)}{d(d+2)(d+4)(d+6)} \left(  \int\frac{d^dp}{(2\pi)^d}\frac{1}{p^4} \right) \Box^4 \Bigg\} \pi(x).
\nonumber \\
&~&
\label{s1tot} \end{eqnarray}
As we have already mentioned, the momentum integrals in the above expressions are assumed to be evaluated with a UV cutoff, as well as 
an IR cutoff of the order of the external momenta in the two-point correlation function. This assumption justifies the  
expansion of the integrant of eq.~(\ref{c22}) in powers of $k$. If an alternative regularization method, such as dimensional
regularization, is employed, the $p$-integral in eq.~(\ref{c22})  must be evaluated without expanding the denominator. 
For $d=4$ the form of eq.~(\ref{s1tot}) is in agreement with the general expectation (\ref{schem}), which was derived under the
assumption $\nu_0\sim 1/\Lx^3$. The unrenormalized effective action includes terms 
$\pi\Box^2\pi$, $\pi\Box^3\pi$ and $\pi\Box^4\pi$, with coefficients that display
quartic, quadratic and logarithmic UV divergences, respectively. If dimensional regularization near $d=4$ was used in 
eq.~(\ref{c22}), the first two terms of eq.~(\ref{s1tot}) would not appear.  
The UV divergence of the third term would correspond to a $1/\ex$ divergence in dimensional regularization. 
The coefficient of the required counterterm, computed in ref.~\cite{quantum2}, agrees with the coefficient of the 
last term in eq. (\ref{s1tot}) for $d=4$.

Our result (\ref{s1tot}) displays the dependence of the various terms on the dimensionality of space. 
Particularly interesting is the absence of
the last term for $d=2$. This feature can be understood as follows: The  operator $K+\Sigma_1+\Sigma_2$ can be written in the
form $K+\Sigma_1+\Sigma_2=-G^{\mu\nu} \partial_\mu \partial_\nu$ with a ``metric"
\be 
G^{\mu \nu}=\eta^{\mu \nu}-2\nu_0\Box \pi\, \eta^{\mu \nu}+2 \nu_0 \partial^\mu \partial^\nu \pi.
\label{metric}
\ee
The operator $G^{\mu\nu} \partial_\mu \partial_\nu$ can be mapped to a similar operator (a covariant Laplacian) 
with the derivatives replaced by covariant derivatives
with Riemann and gauge parts, as explained in section 2 of ref. \cite{vassilevich}. 
In this way our problem can be reduced to the evaluation of $\log {\rm det}( -G^{\mu\nu}\partial_\mu \partial_\nu)$ (the one-loop effective action)
on a ``gravitational" and ``gauge-field" background, constructed from functions of $\pi$ and its derivatives.
\footnote{We emphasize that this mapping and the corresponding backgrounds are not related to the geometric picture of the
brane theory of section \ref{brane}.}
The most efficient method to perform this calculation is through heat kernel techniques \cite{codello2}.
For a gravitational  background in $d=2$, 
one expects a quadratic divergence associated with a cosmological-constant term $\sim \sqrt{G}$.
When the metric is expressed through eq. (\ref{metric}), the resulting term quadratic in $\pi$ is of the form
$\sim \pi \Box^2 \pi$, as shown in ref. \cite{brouz}.
A logarithmic divergence is associated with the 
 Einstein term $\sim \sqrt{G} R$ \cite{codello2}. When expressed in terms of $\pi$, this term is expected to be of the 
form  $\sim \pi \Box^3 \pi$. 
Finite contributions are encapsulated  
by the well known Polyakov action \cite{polyakov}, which is {\it nonlocal}. This implies that 
no local term $\sim \pi \Box^4 \pi$ can be generated in $d=2$. Terms associated with a gauge field background 
display similar behavior.  These considerations provide a plausible explanation why 
a term $\sim \pi \Box^4 \pi$ is not generated by the fluctuation determinant of the operator we are considering.

\section{Quantum corrections in the brane theory}\label{brane}

Our next aim is to examine the quantum corrections in a setting more general than the simple cubic Galileon theory, allowing
for higher-order couplings. This task is complicated by the multitude of possible terms in the action. 
For this reason, we focus on a theory that describes a brane embedded in a flat bulk with one extra dimension. 
In the static gauge, the position modulus of the brane becomes a field of the worldvolume theory, whose 
effective action is strongly constrained by the geometric origin of the construction. In particular, the
allowed terms in the action must respect the bulk Poincar\'e symmetry and 
the reparametrization invariance on the brane \cite{dbigal,multifield}. This constraint limits the 
number of invariants and makes the calculation feasible. Moreover, there is a strong connection between the 
brane and Galileon theories, as has been demonstrated in \cite{dbigal}: The Galileon theory can be obtained from the
brane theory in the nonrelativistic limit $(\partial \pi)^2 \ll 1$, where $\pi$ stands for the brane modulus. 
It must be noted, however, that some invariants of the brane theory must be excluded if only the terms of
the Galileon theory are to be generated in this limit.

We consider a $d$-dimensional brane embedded in a $(d+1)$-dimensional bulk.  
The induced metric on the brane in the static gauge is $g_{\mu\nu}=\eta_{\mu\nu}+\partial_\mu \pi \, \partial_\nu \pi$, where
$\pi$ denotes the extra coordinate of the bulk space. 
We preserve the notation $\eta_{\mu\nu}$
even though we use imaginary time and the bulk metric is Euclidean. 
The extrinsic curvature is 
$K_{\mu\nu}=-\partial_\mu\partial_\nu\pi/\sqrt{1+(\partial\pi)^2}$ and its trace is denoted by $K$. Indices are raised
with the full induced metric. 
The action can be expanded in terms of invariants constructed from the induced metric, the extrinsic curvature and 
covariant derivatives \cite{dbigal,multifield,gliozzi}. 
The leading terms in a curvature expansion are
\begin{eqnarray}
S_\mu&=&\mu\int d^dx  \sqrt{g}=\mu\int d^d x \sqrt{1+(\partial \pi)^2}
\label{sl} \\
S_\nu&=&\nu\int d^dx  \sqrt{g}\, K=-\nu\int d^dx\, \left([\Pi]-\gamma^2[\phi]\right)
\label{sn} \\
S_\kx&=&\frac{\kx}{2}\int d^dx  \sqrt{g}\, K^2=\frac{\kx}{2} \int d^dx\,\gamma \left([\Pi]-\gamma^2[\phi]\right)^2.
\label{sk} \\
S_\kb&=&\frac{\kb}{2}\int d^dx  \sqrt{g}\, R=\frac{\kb}{2}\int d^dx \, \gamma
\left([\Pi]^2-[\Pi^2] +2\gamma^2([\phi^2]-[\Pi][\phi]) \right),
\label{skb} 
\end{eqnarray}
where $\gamma=1/\sqrt{g}=1/\sqrt{1+(\partial \pi)^2}$.
We use the notation of ref.~\cite{dbigal}, with $\Pi_{\mu\nu}=\partial_\mu\partial_\nu \pi$ and square brackets
representing the trace (with respect to $\eta_{\mu\nu}$) of a tensor. Also, we denote 
$[\phi^n]\equiv\partial \pi\cdot \Pi^n \cdot \partial\pi$, 
so that $[\phi]= \partial^\mu\pi \, \partial_\mu\partial_\nu\pi\, \partial^\nu\pi$. 
All dimensionful quantities are expressed in terms of the fundamental scale $\Lx$ of the theory, which is effectively set equal to 1. 
When $d=4$ the couplings $\mu$, $\nu$, $\kb$ correspond to the effective four-dimensional cosmological constant, the five-dimensional
Planck scale $M_5^3$ and the four-dimensional Planck scale $M_4^2$, respectively.

The theory described by the terms (\ref{sl}), (\ref{sn}), (\ref{skb}) belongs to the class of Horndeski
\cite{horndeski} or generalized Galileon theories, which have equations of motion 
that do not involve higher-than-second derivatives of the field $\pi$.  
In particular, the first three terms in the Galileon theory (apart from the tadpole) can be obtained by taking the
nonrelativistic limit $(\partial \pi)^2 \ll 1$ in (\ref{sl}), (\ref{sn}), (\ref{skb}). The term (\ref{sk}) is not included
in the Horndeski class, as it generates higher derivatives in the equation of motion.  
However, the first Gauss-Codazzi equation gives $R=K^2-K^{\mu\nu}K_{\mu\nu}$, which makes it apparent that both terms
(\ref{sk}), (\ref{skb}) must be included at this level of truncation of the brane effective action. 
In the nonrelativistic limit, the term (\ref{sk}) generates
contributions not included in the Galileon theory:
\be
S_\kx=\frac{\kx}{2}\int d^dx \, \pi\Box^2\pi +{\cal O}\left(\pi^4\right).
\label{s0p}
\ee
Since the brane theory involves only field derivatives, the nonrelativistic limit is equivalent to an expansion
in powers of $\pi$.
Up to  (and including)
terms of third order in $\pi$, the brane theory is described by a tree-level action that includes the 
terms of eqs.~(\ref{cubic}) and (\ref{s0p}).

Our interest lies in examining the effect of quantum corrections on the structure of generalized Galileon theories. 
As an interesting example, 
we consider the tree-level brane theory, with action $S_{b0}$ obtained by setting $\mu=1$, $\nu=\nu_0$, $\kx=\kb=0$
in eqs.~(\ref{sl})-(\ref{skb}). 
The one-loop correction is
\be
S_{b1}=\frac{1}{2}{\rm tr}  \log \left(S_{b0}^{(2)} \right).
\label{oneloop} \ee
The calculation is very similar to the one performed in refs.~\cite{ctz,codello} for the $\beta$-functions of the brane 
theory in various dimensions $d$.  
In order to calculate the trace in the
rhs of eq.~(\ref{oneloop}) we need the second functional derivative of the tree-level action on an arbitrary background.  
As explained in the appendix, one finds 
\be
S_{b0}^{(2)}=\Delta +\nu_0V^{\mu\nu}\nabla_\mu \nabla_\nu
+ U+{\cal O}(KR,K^3)\,,
\label{hes}
\ee
where the covariant derivatives are evaluated with the full induced metric, 
$\Delta=-g^{\mu\nu} \nabla_\mu\nabla_\nu$, 
$V^{\mu\nu}= 2(K^{\mu\nu}-K g^{\mu\nu})$ and $U=K^2-K^{\mu\nu} K_{\mu\nu}=R$.
%
%
We substitute the above expression in the rhs of eq.~(\ref{oneloop}) and expand in powers of the curvatures:
\begin{eqnarray}
S_{b1}&=&\frac{1}{2}{\rm tr} \log(\Delta)
+\frac{1}{2}\nu_0{\rm tr}  \left(\frac{1}{\Delta}V^{\mu\nu}\nabla_\mu\nabla_\nu \right)
+\frac{1}{2}{\rm tr}  \left(\frac{1}{\Delta}U\right)\nonumber\\
&&-\frac{1}{4}\nu_0^2{\rm tr}  \left(\frac{1}{\Delta}V^{\mu\nu}\nabla_\mu\nabla_\nu\frac{1}{\Delta} V^{\alpha\beta}\nabla_\alpha\nabla_\beta\right)
+{\cal O}(KR,K^3)\, .
\label{tr}
\end{eqnarray}
The traces in (\ref{tr}) can be computed through the heat kernel expansion, following ref.~\cite{rgmachine}. 
Up to terms of order $K^2$ or $R$ we find
\begin{eqnarray}
&&{\rm tr} \log(\Delta) = \left( \int \frac{d^dp}{(2\pi)^d}\log  p^2 \right) \int d^dx \sqrt{g} 
\nonumber \\
&&
~~~~~~~~~~~~~~~~~~~~~~~~~~~~~~~~~~~~~~~~~~~~~
+\frac{d-2}{12} \left( \int \frac{d^dp}{(2\pi)^d} \frac{\log  p^2 }{p^2}\right)  \int d^dx \sqrt{g}R 
\label{tr2a} \\
&&{\rm tr}  \left(\frac{1}{\Delta}V^{\mu\nu}\nabla_\mu\nabla_\nu \right) = \frac{2(d-1)}{d} \left( \int \frac{d^dp}{(2\pi)^d}
\right) \int d^dx \sqrt{g}K \label{tr2b}\\
&&{\rm tr}  \left(\frac{1}{\Delta}U\right)= \left( \int \frac{d^dp}{(2\pi)^d} \frac{1}{p^2}\right)  \int d^dx \sqrt{g}R \label{tr2c}\\
&&{\rm tr}  \left(\frac{1}{\Delta}V^{\mu\nu}\nabla_\mu\nabla_\nu\frac{1}{\Delta} V^{\alpha\beta}\nabla_\alpha\nabla_\beta\right) = \frac{4(d^2-1)}{d(d+2)} \left(  \int \frac{d^dp}{(2\pi)^d}\right) \int d^dx \sqrt{g}K^2
\nonumber \\
&&
~~~~~~~~~~~~~~~~~~~~~~~~~~~~~~~~~~~~~~~~~~~~~
-\frac{8}{d(d+2)} \left(  \int \frac{d^dp}{(2\pi)^d} \right) \int d^dx \sqrt{g}R\,.
\label{tr2d}
\end{eqnarray}
In equation (\ref{tr2b}) we used $\nabla_\mu \nabla_\nu \rightarrow -\frac{1}{d}g_{\mu\nu} \Delta$, while in equation (\ref{tr2d}) we used the covariant version of eq. (\ref{perm4}). These substitutions are permissible since we are retaining terms up to order $R$ and $K^2$.

Similary to the cubic Galileon results, the momentum integrals in the above expressions are assumed to be evaluated with
UV and IR cutoffs. As the calculation is based on the asymptotic expansion of the heat kernel, the IR cutoff is
taken to be of the order of the typical scale of the curvatures. 

The various terms that appear in eqs.~(\ref{tr2a})-(\ref{tr2d}) involve the curvature invariants
of the effective brane action of eqs.~(\ref{sl})-(\ref{skb}).
Including the tree-level terms,
we obtain the couplings of the theory at one-loop level:
\begin{eqnarray}
\mu&=&1+\frac{1}{2}\int \frac{d^dp}{(2\pi)^d}\log   p^2 
\label{bmuk} \\
\nu&=&\nu_0+\frac{d-1}{d}\nu_0\, \int \frac{d^dp}{(2\pi)^d}
\label{bnuk} \\
\kx&=&-\frac{2(d^2-1)}{d(d+2)} \nu_0^2\, \int \frac{d^dp}{(2\pi)^d}
\label{bkxk} \\
\kb&=&
\frac{4}{d(d+2)} \nu_0^2 \int \frac{d^dp}{(2\pi)^d}
+\int \frac{d^dp}{(2\pi)^d}\frac{1}{p^2}
+\frac{d-2}{12}\int \frac{d^dp}{(2\pi)^d}\frac{\log  p^2 }{p^2}.
\label{bkbk}
\end{eqnarray} 
Parts of the above expressions can be checked through comparison with known results. 
It is apparent from eq.~(\ref{sl}) that the parameter $\mu$ determines the vacuum energy of the theory.
The relation (\ref{bmuk}) contains the correct one-loop contribution arising from the quantum fluctuations of a single 
massless mode in $d$ dimensions. A novel result is obtained if the square root in eq.~(\ref{sl}) 
is expanded in powers of $\pi$. A canonical kinetic term is generated for the field, which receives a wavefunction 
renormalization with a quartic divergence, as given by eq.~(\ref{bmuk}). This very strong effect is a consequence of 
the higher-order derivative interactions obtained in the expansion of the square root at tree level. Reproducing this
result through standard perturbation theory is highly nontrivial and we shall not attempt it here. On the other hand, 
it is clear that the preservation of reparametrization invariance at the quantum level enforces the term of
eq.~(\ref{sl}) to be renormalized maintaining its reparametrization-invariant form. Heat kernel techniques are much
more efficient in realizing this constraint than standard perturbation theory. 

The one-loop correction to $\kx$, given by eq.~(\ref{bkxk}), can be compared with the corresponding one in the cubic Galileon theory. 
In the nonrelativistic limit, the term (\ref{sk}) in the brane action is reduced to  (\ref{s0p}). Substituting in this expression 
the one-loop correction to $\kx$ reproduces exactly the first term of the effective action (\ref{s1tot})
of the cubic Galileon. 
The higher-order derivative interactions of the brane theory 
do not contribute to the renormalization of the operator $\pi\Box^2\pi$. This happens because
the brane and Galileon theories coincide up to the cubic order in an expansion in powers of $\pi$.

\section{Conclusions}\label{conclusions}

The results of section \ref{cubicgalileon} confirmed the expectation for the form of the quantum corrections 
in the cubic Galileon theory, given schematically by eq.~(\ref{schem}). 
The terms quadratic in the field, arising at one-loop level, are given by eq.~(\ref{s1tot}), in which a cutoff regularization in assumed. 
The presence of quartic, quadratic and logarithmic divergences is evident in this expression. The form of the one-loop
effective action confirms that 
the couplings of the cubic Galileon theory do not receive any corrections, and, therefore, are not renormalized.
On the other hand, the quantum corrections generate new terms, not included in the Galileon theory, which would
result in the field equation of motion becoming of higher-than-second order. As a consequence, it is possible that 
the quantum-corrected theory suffers from the presence of ghosts. 

It must be noted that higher-derivative terms are expected to appear even in renormalizable theories when quantum 
corrections are taken into account. However, such terms are not associated with the presence of ghosts.
The physical degrees of freedom are contained in the tree-level action and the higher-order terms 
are interpreted as induced nontrivial interactions. The theory we studied does not fall within this framework.
The quantum corrections imply the presence of a lowest-order term which would be of second order in the field and
fourth order in derivatives. Such a term should have been included already in the tree-level action (\ref{cubic}). 
Its absence requires an extreme fine-tuning. 
If the term is included, the tree-level theory has an equation of motion
of fourth order, which indicates the presence of ghosts as physical degrees of freedom. In the context of Galileon
theories and in the region where the Vainshtein
mechanism operates, 
terms such as the ones appearing in eq. (\ref{schem}) are suppressed relative to the terms in the action (\ref{cubic}),  
 because the background filed is large but has small derivatives. However, the theory still
requires fine-tuning in order to be well defined around the trivial vacuum and not include ghosts in the spectrum.

The brane theory studied in section \ref{brane} belongs to the class of Horndeski or generalized Galileon theories. 
The field equation of motion is of second order, but the theory is not invariant under the Galilean transformation  
$\pi(x) \to \pi(x)+b_\mu x^\mu+c$.
The general form of the quantum corrections in such theories is very complicated because of the large 
number of possible invariants in the action. However, the structure of the brane theory is strongly constrained by its
geometric origin, so that there are actually fewer invariants. Also, the theory is reduced to the Galileon theory in the
nonrelativistic limit and direct comparisons are possible. The study of quantum corrections has revealed 
the emergence of terms not included in the tree-level action, as they result in an equation of motion of higher-than-second order. 
This is a property shared with the cubic Galileon theory. A new feature, not encountered in the Galileon theory, is that 
the tree-level couplings are renormalized. This is apparent in eqs.~(\ref{bmuk}), (\ref{bnuk}) for the couplings $\mu$, $\nu$.
In the nonrelativistic limit, the corresponding terms (\ref{sl}), (\ref{sn}) in the action are reduced
to the standard kinetic and cubic terms of the Galileon theory. As discussed in section \ref{cubicgalileon},
such terms are not renormalized within that theory. We can draw the conclusion that the structure of the generalized 
Galileon theories is altered by quantum corrections more radically than that of the Galileon theory.

Quantum corrections in generalized Galileon theories have also been discussed in the appendix of ref. \cite{quantum2}, where the 
emergence of a larger number of new terms in the action has been observed. On the other, our main conclusion here is the renormalization of 
the couplings of the tree-level action,  a feature differentiating the generalized Galileon theories from the Galileon theory.

As a final remark, we point out that the quantum corrections of the brane theory can also be studied beyond
one-loop perturbation theory. A calculation of the renormalization-group evolution of the couplings within the 
Wilsonian approach has been presented in ref.~\cite{ctz}. The one-loop expressions (\ref{bmuk})-(\ref{bkbk}) can
be obtained at the first level of an iterative solution of the evolution equations. Moreover, new properties of
the theory can be investigated, such as the possible presence of a UV fixed point that could  underlie the UV completion
of the theory.

\section*{Acknowledgments}
The work of N.T. has been supported in part 
by the European Commission under the ERC Advanced Grant BSMOXFORD 228169.
It has also been co-financed by the European Union (European Social Fund – ESF) and Greek national 
funds through the Operational Program ``Education and Lifelong Learning" of the National Strategic Reference 
Framework (NSRF) - Research Funding Program: ``THALIS. Investing in the society of knowledge through the 
European Social Fund".
The research of O.Z. is supported by the DFG within the Emmy-Noether program (Grant SA/1975 1-1). 

\section*{Appendix}

A $d$--dimensional brane embedded in $\mathbb{R}^{d+1}$ is a
function $\mathbf{r}(x):\mathbb{R}^{d}\rightarrow\mathbb{R}^{d+1}$
which maps the point $x$ in $\mathbb{R}^{d}$ to the point $\mathbf{r}$
in $\mathbb{R}^{d+1}$. At every point on the surface the tangent
vectors $\mathbf{r}_{\alpha}=\partial_{\alpha}\mathbf{r}$, $\alpha=1,...,d$
and the normal unit vector $\mathbf{n}$ form a basis of $\mathbb{R}^{d+1}$.
We consider variations of $\mathbf{r}$ in the normal direction, because those in the tangential directions correspond to
reparametrizations that can be eliminated in the appropriate gauge. We define
\[
\delta\mathbf{r}= \nu \, \mathbf{n}.
\]
The induced metric tensor (first fundamental form) is defined as
\[
g_{\alpha\beta}=\mathbf{r}_{\alpha}\cdot\mathbf{r}_{\beta}
\]
and can be used to construct the volume element on the brane $dV=\sqrt{g}dx_{1}\cdots dx_{d}$,
where $g=\det g_{\alpha\beta}$. We associate with the normal direction
the exterior curvature tensor (second fundamental form) defined as
\[
K_{\alpha\beta}=-\mathbf{r}_{\alpha}\cdot\mathbf{n}_{\beta}\,.
\]
The intrinsic scalar curvature is related to the extrinsic curvature by Gauss's Theorema Egregium:
\[
R=K^2-K^{\alpha\beta} K_{\alpha\beta}\,.
\]

The first two reparametrization-invariant quantities we can construct on the brane are
\[
I_{0}=\int d^d x\sqrt{g}\qquad\qquad I_{1}=\int d^d x\sqrt{g}\, K.
\]
They constitute the first two invariants (\ref{sl}), (\ref{sn}) in the brane action of section 3.
In order to compute their variations we take the basic variations of the metric
\begin{eqnarray}
\delta\sqrt{g} = -\sqrt{g}\nu K \qquad\qquad
\delta^{2}\sqrt{g} = \sqrt{g}\left\{ \nabla_{\alpha}\nu\nabla^{\alpha}\nu+\nu^{2}R\right\} \label{app_2}
\end{eqnarray}
and of the trace of the extrinsic curvature
\begin{eqnarray}
\delta K & = & \nabla^{2}\nu+(K^{2}-R)\nu\nonumber \\
\delta^{2}K & = & -K\nabla^{\gamma}\nu\nabla_{\gamma}\nu+2\nu K^{\alpha\beta}\nabla_{\alpha}\nabla_{\beta}\nu+2K^{\alpha\gamma}K_{\gamma}^{\beta}K_{\alpha\beta}\nu^{2}\label{app_4}
\end{eqnarray}
from ref. \cite{Capovilla2003}. Then the first variations are
\begin{equation}
\delta I_{0}=\delta\int d^d x\sqrt{g}=-\int d^d x\sqrt{g}K\nu\label{app_7}
\end{equation}
and
\begin{equation}
\delta I_{1}=\delta\int d^d x\sqrt{g}K=\int d^d x\left[\delta(\sqrt{g}) K+\sqrt{g}\delta( K)\right]
=-\int d^d x\sqrt{g}\left[\Delta+R\right]\nu\,.\label{app_8}
\end{equation}
The second variations of $I_{0}$ follows directly from eq. (\ref{app_8}):
\begin{equation}
\delta^{2}I_{0}=-\delta\int d^d x\sqrt{g}K\nu=\int d^d x\sqrt{g}\nu\left[\Delta+R\right]\nu\,.\label{app_9}
\end{equation}
We need the second variation of $I_{1}$ only to second order in the
curvature:
\begin{eqnarray}
\delta^{2}I_{1} & = & \int d^d x\left[\delta^{2}(\sqrt{g})\, K+2\delta(\sqrt{g})\, \delta (K)+\sqrt{g}\, \delta^{2}(K)\right]
\nonumber \\
 & = & \int d^d x\sqrt{g}\left\{ K\nabla_{\alpha}\nu\nabla^{\alpha}\nu+2\nu K\Delta\nu\right.
\left.-K\nabla^{\gamma}\nu\nabla_{\gamma}\nu+2\nu K^{\alpha\beta}\nabla_{\alpha}\nabla_{\beta}\nu\right\} +O(KR,K^{3})\nonumber \\
 & = & 2\int d^d x\sqrt{g}\left(K^{\alpha\beta}-g^{\alpha\beta}K\right)\nu\nabla_{\alpha}\nabla_{\beta}\nu+O(KR,K^{3})\,.\label{app_10}
\end{eqnarray}
From eqs. (\ref{app_9}) and (\ref{app_10}) we can read off the differential operator (\ref{hes}) of section 3.


\begin{thebibliography}{999}

\bibitem{dgp}
  G.~R.~Dvali, G.~Gabadadze and M.~Porrati,
  Phys.\ Lett.\ B {\bf 485} (2000) 208  [hep-th/0005016];
\\
  C.~Deffayet, G.~R.~Dvali and G.~Gabadadze,
  Phys.\ Rev.\ D {\bf 65} (2002) 044023  [astro-ph/0105068].  

\bibitem{galileon}
  A.~Nicolis, R.~Rattazzi and E.~Trincherini,
  Phys.\ Rev.\ D {\bf 79} (2009) 064036  [arXiv:0811.2197 [hep-th]].  

\bibitem{horndeski}
  G.~Horndeski,
  Int.\ J.\ Theor.\ Phys. {\bf 10} (1974) 363.

\bibitem{genegal}
  C.~Deffayet, X.~Gao, D.~A.~Steer and G.~Zahariade,
  Phys.\ Rev.\ D {\bf 84} (2011) 064039  [arXiv:1103.3260 [hep-th]];
\\
  T.~Kobayashi, M.~Yamaguchi and J.~'i.~Yokoyama,
  Prog.\ Theor.\ Phys.\  {\bf 126} (2011) 511  [arXiv:1105.5723 [hep-th]];
\\
  C.~Charmousis, E.~J.~Copeland, A.~Padilla and P.~M.~Saffin,
  Phys.\ Rev.\ Lett.\  {\bf 108} (2012) 051101  [arXiv:1106.2000 [hep-th]].  

\bibitem{quantum1}
 A.~Nicolis and R.~Rattazzi,
 JHEP {\bf 0406} (2004) 059  [hep-th/0404159].

\bibitem{nonrenorm}
  C.~de Rham, G.~Gabadadze, L.~Heisenberg and D.~Pirtskhalava,
Phys.\ Rev.\ D {\bf 87} (2013) 085017  [arXiv:1212.4128];
\\
  C.~de Rham, L.~Heisenberg and R.~H.~Ribeiro,
  Phys.\ Rev.\ D {\bf 88} (2013) 084058
  [arXiv:1307.7169 [hep-th]];
\\
  C.~de Rham, G.~Gabadadze, L.~Heisenberg and D.~Pirtskhalava,
  Phys.\ Rev.\ D {\bf 83} (2011) 103516
  [arXiv:1010.1780 [hep-th]].

\bibitem{multifield}
 K.~Hinterbichler, M.~Trodden and D.~Wesley,
Phys.\ Rev.\ D {\bf 82} (2010) 124018  [arXiv:1008.1305 [hep-th]].

\bibitem{vainshtein}
  A.~I.~Vainshtein,
 Phys.\ Lett.\ B {\bf 39} (1972) 393.  

\bibitem{quantum2}
  T.~de Paula Netto and I.~L.~Shapiro,
Phys.\ Lett.\ B {\bf 716} (2012) 454  [arXiv:1207.0534 [hep-th]].  

\bibitem{polyakov}
 A.~M.~Polyakov,
  Phys.\ Lett.\ B {\bf 103} (1981) 207.

\bibitem{vassilevich}
  D.~V.~Vassilevich,
  Phys.\ Rept.\  {\bf 388} (2003) 279
  [hep-th/0306138].

\bibitem{codello2}
  A.~Codello,
  Annals Phys.\  {\bf 325} (2010) 1727
  [arXiv:1004.2171 [hep-th]].

\bibitem{brouz}
  N.~Brouzakis and N.~Tetradis,
  arXiv:1401.2775 [hep-th].

\bibitem{dbigal}
  C.~de Rham and A.~J.~Tolley,
  JCAP {\bf 1005} (2010) 015  [arXiv:1003.5917 [hep-th]].  

\bibitem{gliozzi}
 O.~Aharony and M.~Field,
JHEP {\bf 1101} (2011) 065  [arXiv:1008.2636 [hep-th]];  
\\
  O.~Aharony and M.~Dodelson,
JHEP {\bf 1202} (2012) 008  [arXiv:1111.5758 [hep-th]];  
\\
  F.~Gliozzi and M.~Meineri,
JHEP {\bf 1208} (2012) 056  [arXiv:1207.2912 [hep-th]].  

\bibitem{ctz}
  A.~Codello, N.~Tetradis and O.~Zanusso,
JHEP {\bf 1304} (2013) 036  [arXiv:1212.4073 [hep-th]].  


\bibitem{codello}
  A.~Codello and O.~Zanusso,
  Phys.\ Rev.\ D {\bf 83} (2011) 125021  [arXiv:1103.1089 [hep-th]].  

\bibitem{rgmachine}
  D.~Benedetti, K.~Groh, P.~F.~Machado and F.~Saueressig,
  JHEP {\bf 1106} (2011) 079
  [arXiv:1012.3081 [hep-th]];
\\
  K.~Groh, F.~Saueressig and O.~Zanusso,
  [arXiv:1112.4856 [math-ph]].
  
\bibitem{Capovilla2003}R.~Capovilla,  J.~Guven and J.~A.~Santiago,
J. Phys. A: Math. Gen. {\bf 36} (2003) 6281,
[arXiv:cond-mat/0212118v1].  

\end{thebibliography}
\end{document}